\documentclass[
preprint,
amsmath,
amssymb,
aps,
pra,
floatfix,
]{revtex4-2}

\usepackage{graphicx} 
\usepackage{amsmath}
\usepackage{xcolor}
\usepackage{hyperref}

\begin{document}

\title{Quantum many-body effects in the optical response of ideal thin films}

\author{David Trejo-Garcia}
\affiliation{%
Faculty of Engineering and Natural Sciences, Tampere University, Tampere, Finland
}%

\author{Tapio T. Rantala}
\affiliation{%
Faculty of Engineering and Natural Sciences, Tampere University, Tampere, Finland
}%

\author{Marco Ornigotti}
\affiliation{%
Faculty of Engineering and Natural Sciences, Tampere University, Tampere, Finland
}%

\author{Juha Tiihonen}
\email{juha.tiihonen@tuni.fi}
\affiliation{%
Faculty of Engineering and Natural Sciences, Tampere University, Tampere, Finland
}%

\date{\today}

\begin{abstract}
We study quantum many-body effects in the long-wavelength optical response of confined electrons at finite temperatures. We simulate homogeneous electron gas confined in one dimension into a slab of nanoscale thickness. We demonstrate how the slab boundaries break down the ideal Drude response of free charge carriers, giving rise to scattering effects due to both the surfaces and quantum many-body interactions. We use a recent path-integral Monte Carlo (PIMC) approach developed in [Tiihonen et al. Phys. Rev. A 113, 053711] to quantify these effects in high accuracy. We perform phenomenological fits to Drude and Drude--Lorentz models parameters, manifesting various trends of the optical response with physical parameters like density and temperature, and numerical effects like finite size and the quantum statistics.
\end{abstract}

\maketitle

\section{Introduction}
\label{sec:introduction}

In recent years, material structures in reduced dimensions and sizes have gained intrigue due to the tunability of properties by geometrical design. For instance, since the discovery of graphene, thin 2D films of semiconductors have been studied for their enhanced and anomalous optical properties \cite{Chhowalla2016, Bandurin2016}. Confinement effects are interesting in other dimensionalities, like surface plasmon resonances showing in 1D nanorods \cite{Baida2011} and finite metal clusters \cite{Ekardt1984, Chaudhary2024}. The potential of nanoconfinements for new discoveries and technologies is rich beyond measure, and to match the rapid development, techniques both experimental \cite{saini_near_2022, Mekhael2024} and numerical are striving to keep up.

The optical response of atomic-scale structures must be treated using quantum physics. The state-of-the-art numerical approaches include the density function theory (DFT) or the time-dependent DFT (TDDFT) \cite{Yabana1996}, which recover optical properties from first principles, including electrons interacting with realistic ion lattices and one another through exchange--correlation (XC) functionals. The atomistic spectra with various scattering channels (like interband, intraband, electron--ion) offer a sophisticated reference beyond simplified models like the electron gas \cite{Yabana1996, Povarnitsyn2012}. At finite temperatures, the DFT optical response can be obtained from the Kubo--Greenwood (KG) formalism, which is amenable to molecular dynamics (MD) simulation of ionic thermodynamics \cite{Dufty2018, Melton2024}. However, reaching high temperatures within the KG formalism becomes troublesome as more and more single-particle orbitals are activated \cite{Melton2024}. While the performance of DFT is often satisfactory, its contextual validation and systematic improvement remain hard without extrinsic references. To inform the development of scalable methods, like DFT, for fundamental \cite{Ceperley1980, Brown2013} and optical response properties \cite{Ramakrishna2021}, accurate benchmark data from Monte Carlo (MC) methods can be vital \cite{Giuliani2008}.

In this work, we apply a path-integral Monte Carlo (PIMC) approach that was recently developed to study the optical (long-wavelength) response of quantum plasmas \cite{Tiihonen2026}. The method complements mean-field methods and the KG formalism in that it high temperatures and numerically exact quantum interactions gracefully. The PIMC importance sampling is straightforward and robust, avoiding typical difficulties of integrating in semi-infinite systems \cite{Mazzei2022}, but it faces truncation effects due to the finite size \cite{Fraser1996}. In this work, we focus on the non-periodic properties, where the finite-size effects are less imminent. In this domain, our real-space formulation of the optical response \cite{Tiihonen2016a, Tiihonen2018, Tiihonen2019} is also more natural than the commonly used structure factor \cite{Giuliani2008, Hamann2020, Dornheim2023}. On the other hand, PIMC suffers challenges due to the Fermion sign problem \cite{Loh1990, Ceperley1995} (FSP) and the lack of straightforward real-time observables. Because of the FSP, the simulations are limited to relatively small sizes and high temperatures. Model potentials must be used instead of atomistic simulations. More performant PIMC sampling approaches exist \cite{Boninsegni2006, Dornheim2015, Xiong2022, Dornheim2023b} that could be considered in future works. Most notably, however, the restricted-PIMC method \cite{Ceperley1996} cannot be used to study non-commuting observables, such as the optical response.

We consider the homogeneous electron gas (HEG) confined in one dimension into a quasi-2D slab geometry. The optical response of the confined system is compared to the ideal gas, whose long-wavelength response, the Drude susceptibility, is agnostic of quantum many-body interactions because of perfect screening \cite{Nozieres1999}. When the translation symmetry is broken by the confinement, traces of the interactions also emerge. We survey, on an effective level, how these effects depend on the relevant parameters, such as the confinement size, finite temperature, particle density. We also study effects due to the particle number and quantum statistics. The analysis is mainly done in the imaginary domain of the PIMC observables, using Drude and Drude--Lorentz models for quantification and analytic continuation. In practice, we recover finite scattering rates $\Gamma$ and resonances $\omega_0$ in regimes, where the confinement size is proportional to lengths occupied by few quantum particles. Direct comparison with, \textit{e.g.}, DFT would be highly interesting but also obscured by the inconsistencies mentioned earlier. In this work we focus on a proof of principle, drawing from the intuitive appeal of the PIMC methodology, and leave the comparison with DFT for future works.

The rest of the work is organized as follows: In Sec.~\ref{sec:theory}, we review the key method developments, including the estimator and the model potentials. In Sec.~\ref{sec:computational_details}, we detail the numerical implementations. In Sec.~\ref{sec:results}, we discuss the results, and in Sec.~\ref{sec:summary} we summarize the work.

\subsection{Theory}
\label{sec:theory}

Here, we first review the optical response framework in the complex domain, as laid out in Ref.~\cite{Tiihonen2026}. Next, we review details of the PIMC simulation and the estimator. Finally, we define different model potentials to be used in simulations. For convenience, the formulas are presented in Hartree atomic units, i.e.,  $\hbar = 4 \pi \varepsilon_0 = m_e = 1$.

\subsection{Optical response in imaginary time}

The long-wavelength dielectric function of a quantum plasma is given by
\begin{align}
    \varepsilon(\omega) &= 1 + \chi(\omega)
\end{align}
with optical susceptibility
\begin{align}
    \label{eq:chi_w}
    \chi(\omega) &= -4 \pi i n G^R(\omega),
\end{align}
where $n$ is the particle density and
\begin{align}
    G^R(\omega) &= \mathcal{F}^{-1} G^R(t) \\
    \label{eq:Gt}
    G^R(t) &= \Theta(t) \langle \mu(t) \mu(0) \rangle
\end{align}
is the retarded dipole autocorrelation function, $\Theta$ the Heaviside function, $\mathcal{F}^{-1}$ the inverse Fourier transform and $\mu(t)$ the dipole moment at time $t$ after incidence. To solve the complex $G^R(\omega)$ is to solve the dielectric function.

The correlation function $G^R(\omega)$ is analytic in the upper complex plane and has the following non-negative spectral function:
\begin{equation}
    \label{eq:spectrum}
    A(\omega) \equiv -2 \mathrm{Im}[G^R(\omega)].
\end{equation}
The spectral density allows analytic continuation of $G^R$ to imaginary time $-it \rightarrow \tau$ in either time or frequency domain through \cite{Jarrell1996, Tiihonen2026}
\begin{align}
    \label{eq:Aw_to_Gtau}
    \mathcal{G}(\tau) = \int_{-\infty}^{\infty} \frac{d \omega}{2 \pi} \frac{\mathrm{e}^{-\tau\omega}}{1 - \mathrm{e}^{-\beta \omega}} A(\omega),
\end{align}
where $\beta = 1 / k_B T$ is the inverse temperature. The imaginary-time correlation function $\mathcal{G}(\tau)$ has bosonic symmetry with $\mathcal{G}(\tau) = \mathcal{G}(\tau + n\beta)$ for any integer $n$, and similar to its counterpart $G^R(t)$, is evaluated with 
\begin{equation}
    \label{eq:Gtau}
    \mathcal{G}(\tau) = \frac{1}{\hbar} \langle \mathcal{T} \mu_z(0) \mu_z(\tau) \rangle/N,
\end{equation}
where $\mathcal{T}$ is the time-ordering operator. The dipole moment operators $\mu_z(\tau)$ are also assumed normal ordered, meaning $\langle \mu_z(\tau) \rangle = 0$. The Fourier transform of $\mathcal{G}(\tau)$ is the Matsubara series $\mathcal{G}(i \omega_n)$, where $\omega_n = 2 \pi n / \hbar \beta$ for all integers, which can also be obtained with \cite{Jarrell1996}
\begin{equation}
    \label{eq:Aw_to_Giw}
    \mathcal{G}(i \omega_n) = -\int_{-\infty}^{\infty} \frac{d \omega}{2 \pi} \frac{1}{i\omega_n - \omega} A(\omega).
\end{equation}

\subsection{Complex Drude--Lorentz model}
\label{sec:drude}

The Drude susceptibility for a population of noninteracting particles is given by
\begin{equation}
    \chi^D(\omega) = -\frac{\omega^2_p}{\omega^2 + i\omega\Gamma},
    \label{eq:GD}
\end{equation}
where $\omega_p^2=4 \pi n$ (in a.u.) is the plasma frequency squared and $\Gamma$ is a phenomenological scattering rate. Eq.~\eqref{eq:GD} can be made into the Drude--Lorentz model
\begin{equation}
    \chi^{DL}(\omega) = -\frac{i\omega_p^2}{4 \pi n}\frac{1}{\omega^2 - \omega_0^2 + i\omega\Gamma}
    \label{eq:GDL}
\end{equation}
by introducing a resonance frequency $\omega_0$. Based on Eqs.~\eqref{eq:chi_w} and \eqref{eq:spectrum}, the microscopic dynamical spectra are given (in a.u.) by
\begin{align}
    A^D(\omega) &= \frac{2}{\omega^3/\Gamma + \omega \Gamma} \\
    A^{DL}(\omega) &= \frac{2\omega \Gamma}{(\omega^2 - \omega_0^2)^2 + \omega^2\Gamma^2} \\
\end{align}
and the respective Matsubara spectra are
\begin{align}
    \mathcal{G}^D(i \omega_n) &= \frac{2\pi}{\omega_n^2 + \omega_n \Gamma}, 
    \label{eq:Giw_D} \\
    \mathcal{G}^{DL}(i \omega_n) &= \frac{2\pi}{\omega_n^2 + \omega_n \Gamma  + \omega_0^2}.
    \label{eq:Giw_DL}
\end{align}
Clearly, the Drude--Lorentz ansatz has more expressive power through the resonance frequency $\omega_0$. However, as both terms, $\omega_0$ and $\Gamma$, compete to inversely scale down the values of $\mathcal{G}(i\omega_n)$, their estimation is prone to misidentification and less robust than the non-resonant Drude model.

Eqs.~\eqref{eq:Giw_D} and \eqref{eq:Giw_DL} will be used as the most simple Ans\"atze for performing analytic continuation from Matsubara spectrum back to real domain. This amounts to the inversion of Eq.~\eqref{eq:Aw_to_Giw}, which is a well-known ill-posed problem, when the data has random noise \cite{Jarrell1996}. Yet, many methods have been designed to tackle this problem in generic situations. One of the most popular is the maximum entropy method \cite{Jarrell1996}, but it treats Drude-like divergent spectra poorly. Our simple ansatz can provide coarse insights toward dynamical spectra, and it could be easily extended to contain multiple resonances. However, more sophisticated methods of analytic continuation will be left for future.

\subsection{Path-integral Monte Carlo}

Quantum statistical averages of the dipole autocorrelation function can be estimated with the canonical PIMC method \cite{Ceperley1995} at finite temperatures. For clarity, the expressions in this section are superficial, and more rigorous and generalized presentations are found elsewhere, \textit{e.g.} Ref.~\cite{Dornheim2023b}. In PIMC, one uses the Metropolis algorithm to draw samples from the numerically exact thermal partition function for $N$ particles in the the Boson (B) or Fermion (F) statistics:
\begin{equation}
    \label{eq:Z_perm}
    Z_{B/F} = \frac{1}{N!} \sum_{\mathcal{P}} (\pm1)^{\mathcal{P}} \int \mathrm{d}R \langle R | \hat{\rho}(\beta) | \mathcal{P}R \rangle,
\end{equation}
where $\hat{\rho}(\beta) = \mathrm{e}^{-\beta H_0}$ is the density operator, $H_0$ is the Hamiltonian, and $\mathcal{P}$ is the permutation operator. Since the evaluation of $N!$ permutations grows unfeasible, the Metropolis sampling is done by sampling individual permutations between indistinguishable species. Without the permutations (\textit{i.e.} exchange), the particles are considered distinguishable, fictitiously, and often referred to as \textit{Boltzmannons}.

Getting proper Fermion statistics from statistical sampling requires special care, because the factor $(-1)^\mathcal{P}$ renders the weight of all odd permutations negative, which cannot be treated as a probability. Therefore, the sampling will be done on a modified partition function $Z'$ with strictly positive weights. For this, we use the algorithm described in Ref.~\cite{Boninsegni2005}. Proper observables $O$ are then recovered from
\begin{equation}
    \label{eq:signed_obs}
    \langle O \rangle = \frac{\langle O s \rangle'}{\langle s \rangle'},
\end{equation}
where $\langle s \rangle'$ is the average sign from the modified sampling. The average sign is the mean over instantaneous signs, which are $\pm 1$ depending on the permutation count. Consequently, it also measures the statistical efficiency of the observables. The sign goes lower with $N$, $\beta$ and $n$ as the occurrences of even (positive) and odd (negative) permutations almost completely cancel out, giving rise to the notorious sign problem.

Beside the sign, estimation of energies and observables is similar to Ref.~\cite{Tiihonen2026}. The dipole autocorrelation function can be estimated from
\begin{align}
    \label{eq:relative_estimator}
    \mathcal{\tilde G}(\tau) &\equiv \mathcal{G}(\tau) - \mathcal{G}(0) \\
    &= -\frac{1}{2N} \langle (\mu(0) - \mu(\tau))^2 \rangle,
\end{align}
where the translation invariance of periodic systems is subtracted out at the expense of losing the static polarizability $\mathcal{G}(0)$. For more information, see discussions in Ref.~\cite{Tiihonen2026}. While it would be possible to calculate $\mathcal{G}(\tau)$ in confined systems, we will keep to Eq.~\eqref{eq:relative_estimator} and focus on dynamical properties. Furthermore, the winding constraint described in Ref.~\cite{Tiihonen2026} is implemented differently in this work: Instead of rejecting moves that introduce non-zero periodic winding of the trajectories, we rectify the trial paths before evaluation. The rectification is done by calculating the difference between the right and the wrong way for passing the periodic box and then fixing the wrong way by adding linear interpolation of the difference vector.

\subsection{Slab confinement}
\label{sec:potentials}

To realize the quasi-2D slab confinement, we immerse the particles in an external potential of the form
\begin{align}
    \label{eq:confinement}
    V(\mathbf{r}, R) = \tfrac{1}{2} \Omega_0^2 \Delta_i(\mathbf{r}, R),
\end{align}
where $\Omega_0^2$ is the confinement strength, $R>0$ is a confinement size parameter and $\Delta_i(\mathbf{r}, R)$ is a displacement function. In 1D confinement, it is expressed as
\begin{align}
    \Delta_1(\mathbf{r}, R) = \left\{
    \begin{array}{lr}
        0, & r_1 \leq R \\
        (r_1 - R)^2, & r_1 > R
    \end{array}
    \right.,
\end{align}
where $r_1 = x^2$ is a displacement variable chosen without loss of generality along the $x$ coordinate around $x=0$. In practice, the function defines a zero-potential region bounded by harmonic walls beyond $|x| > R$, while the other coordinates are subject to periodic boundary conditions. The harmonic bounding potential with $\Omega_0^2 R^2 \gg \hbar$ is a simple, phenomenological model to accomplish soft boundaries akin to real surfaces.

For finite-sized simulations, the periodic box sizes $L$ are associated with the number of particles $N$ to meet a given finite particle density as follows:
\begin{equation}
    L^2_y = L^2_z = \frac{2 \pi r_s^3}{3}\frac{N}{R},
\end{equation}
where $2R L_y L_z$ is regarded as the total volume.

\section{Computational details}
\label{sec:computational_details}

Same as in Ref.~\cite{Tiihonen2026}, The PIMC simulation software implements the canonical Metropolis Monte Carlo algorithm for a PIMC walker in Fortran90, featuring matrix squaring of the exact Coulomb pair action, parallel sampling and data binning of Markovian walkers. The simulations are semi-infinite, but the periodic images are not treated explicitly or using Ewald summation techniques \cite{Fraser1996}. The justification is that we focus on properties associated with open boundaries and not, for instance, energies. This choice does intensify finite-size effects, as discussed in Sec.~\ref{sec:results}, and treating periodic images will be a worthy investment in the future. Throughout the simulations, we use a finite time-step $\Delta \tau = 0.001 * r_s^3$, which has been tested out to give a satisfactory performance \cite{Tiihonen2026}.

The simulation campaigns are operated with Nexus workflows \cite{Krogel2016} on various high-performance computing facilities listed in Sec.~\ref{sec:acknowledgements}. Unless mentioned otherwise, all simulations are based on Fermion statistics. However, the Fermion simulations are routinely preceded by a warm-up run without permutations (\textit{i.e.} Boltzmannons) to speed up equilibration without tangling up the walker. The data postprocessing is done using standard Python libraries. More implementation and data processing details can be found in Ref.~\cite{Tiihonen2026} and in a separate data repository \cite{Repository}.

Results of the PIMC simulation, \textit{i.e.}, energies and correlation functions, have finite statistical uncertainties due to finite sampling. The uncertainties are estimated based on 2$\sigma$ standard error of the mean (2SEM), including sample autocorrelation time $\kappa$, as described in Refs.~\cite{Kylaenpaeae2011, Tiihonen2019} with the extension that each of the input data in a $J$-long measurement sequence $\{O_1, \ldots O_J\}$ be multiplied by $s_j / \sum_J s_j$ to straightforwardly factor in the sign. The effective means and $\kappa$ per each observable, including $\mathcal{\tilde G}(\tau)$ for each individual $\tau$, are analyzed from the sequential distribution of sample block averages. The block averages are means over numerous measurements, aggregated in parallel from independent MC walkers and also by binning subsequent measurements from each walker. Within each walker, a number of MC moves is performed between each measurement to decrease sample autocorrelation. Statistical uncertainties of $\mathcal{\tilde G}(i \omega_n)$ are measured from the Fourier transforms of $\mathcal{\tilde G}(\tau)$ block averages. The uncertainties of $\Gamma^D$, $\Gamma^{DL}$ and $\omega_0$ are the $95 \%$ confidence intervals of bootstrap resampled fits.

\section{Results}
\label{sec:results}

We present results based on HEG confined in a slab potential described in Sec.~\ref{sec:potentials}. Because of the perfect screening, the optical response of the ideal HEG equals to the Drude spectrum given in Eq.~\eqref{eq:GD} in the low-scattering limit \cite{Tiihonen2026}, namely
\begin{align}
    \mathcal{\tilde G}_0^D(\tau) &\equiv \mathcal{\tilde G}^D(\tau, \Gamma \rightarrow 0) = \frac{\tau^2 - \tau \beta}{4 \lambda \beta} \\
    \mathcal{\tilde G}_0^D(i \omega_n \neq 0) &\equiv \mathcal{G}^D(i \omega_n \neq 0, \Gamma \rightarrow 0) = \frac{2 \pi}{\omega_n^2} \\
    \mathcal{\tilde G}_0^D(i \omega_n = 0) &\equiv \int_0^\beta \mathrm{d}\tau \mathcal{\tilde G}_0^D(\tau) = \frac{\beta^2}{24\lambda},
\end{align}
where $\lambda=1/2m_e$ and the last part is an nonphysical measure of apparent scale \cite{Tiihonen2026}, not used in the fitting. Thus, our analyses are based on comparing against this baseline upon breaking of the translation symmetry along the $x$ axis. The basic quantity is therefore the relative deviation
\begin{align}
    \Delta \mathcal{\tilde G}(\tau) &= \frac{\mathcal{\tilde G}(\tau) - \mathcal{\tilde G}_0^D(\tau)}{\mathcal{\tilde G}_0^D(\tau)}
    \label{eq:deltaG}.
\end{align}
Furthermore, as already suggested in \cite{Tiihonen2026}, the deviation can be associated with effective Drude scattering. We will therefore use least-squares fitting of the free parameters in the the Drude and Drude-Lorentz models given respectively in Eq.~\eqref{eq:Giw_D} ($\Gamma^D$) and \eqref{eq:Giw_DL} ($\Gamma^{DL}$, $\omega_0$). Different labels will be used to stress the inconsistency of the fits, namely $\Gamma^D$ and $\Gamma^{DL}$ cannot be expected to match when $\omega_0$ is activated. As will become evident, the fitted parameters can be used to gauge scattering effects both due to the confinement boundaries and the particle interactions. The effects trend simultaneously, albeit differently, versus spatial dimensions, quantum statistics, finite temperatures, densities and the simulation size.

The simulation geometries of each confinement are characterized by $r_s$, $R$ and $N$. Furthermore, we express the finite temperatures with $\Theta=T/T_f$, where $k_B T_f= \hbar^2 k_f^2/2$ and $k_f(\xi) = [3 n \pi^2(1 + \xi)]^{1/3}$ are, respectively, the Fermi temperature and the Fermi wavevector, and $\xi$ is the spin polarization \cite{Giuliani2008}. Here, we only consider spin-unpolarized gas $\xi=0$ with $N_\uparrow = N_\downarrow = N/2$. When compared to a \textit{Boltzmannon} simulation with distinguishable particles, the Fermi temperature is also considered that of the unpolarized gas. The total number of particles is varied up to $N \leq 32$ to investigate finite-size effects. All results are presented in Hartree atomic units with $m=m_e=1$ and $e^2=1$. The slab thicknesses $2R$ are varied between $5-120$ Bohr, so from atomically thin to up to 6.4 nanometers.

\begin{figure}[t]
   \centering
   \includegraphics[width=\textwidth]{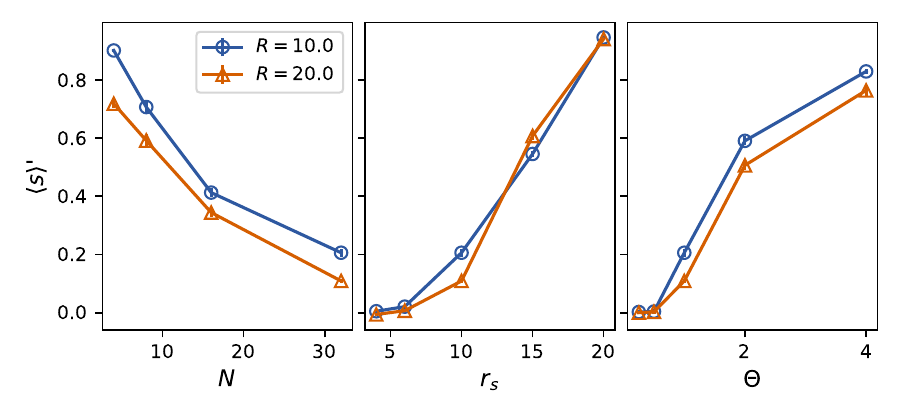}
   \caption{Plots of the average sign $\langle s \rangle'$ of spin-unpolarized simulation versus simulation parameters in slab confinements at two different thicknesses and hardness of $\Omega^2=10$ : (left) sign versus $N$ with $r_s=10$ and $\Theta=1$; (middle) sign versus $r_s$ with $N=32$, $\Theta=1$; (right) sign versus $\Theta$ with $r_s=10$, $N=32$. The lines are drawn to guide the eye.
   }
   \label{fig:sign}
\end{figure}

Because of the FSP, we will only study relatively low densities and high temperatures. To illustrate this, we present in Fig.~\ref{fig:sign} the average sign $\langle s \rangle'$ of the PIMC calculation versus key simulation parameters $r_s$, $\Theta$ and $N$. As expected, the sign dies out toward low $\Theta$ and $r_s$ and high $N$, to the detriment of the numerical efficiency. When compared between thinner (blue) and thicker (orange) slabs, the sign is modestly higher for the thinner one, as it suppresses more the likelihood of permutations. In this proof-of-concept survey, we shall not chase heroic calculations to compensate for the low sign. Rather, we only report results when the sign is reasonable (\textit{i.e.} $\langle s \rangle' \sim 0.05$ or higher). This limits the physical parameters to densities around $2 \ldots 20 \times 10^{20}$ particles per cm$^3$ and temperatures to at least $1500 \ldots 6000$ Kelvin. Corresponding parameters could be found in doped semiconductor thin films like TiO$_2$ \cite{He2023} or indium tin-oxide \cite{Gondorf2011, Blair2023}, except that we use vacuum permittivity. The temperatures of thousands of Kelvin are typical in the thermalization emission of ultrashort laser pulses \cite{Bresson2020}, often considered in the two-temperature model \cite{Anisimov1974}.


\begin{figure}[h]
   \centering
   \includegraphics[width=6.0cm]{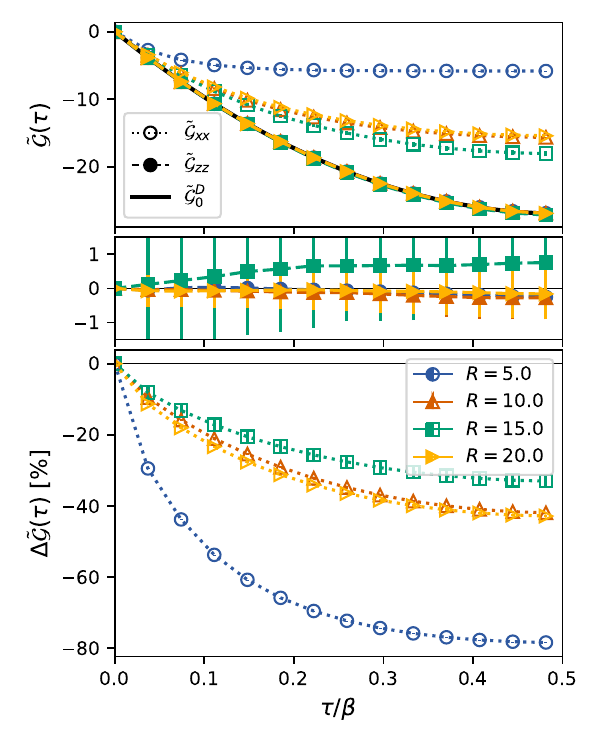}
   \includegraphics[width=6.0cm]{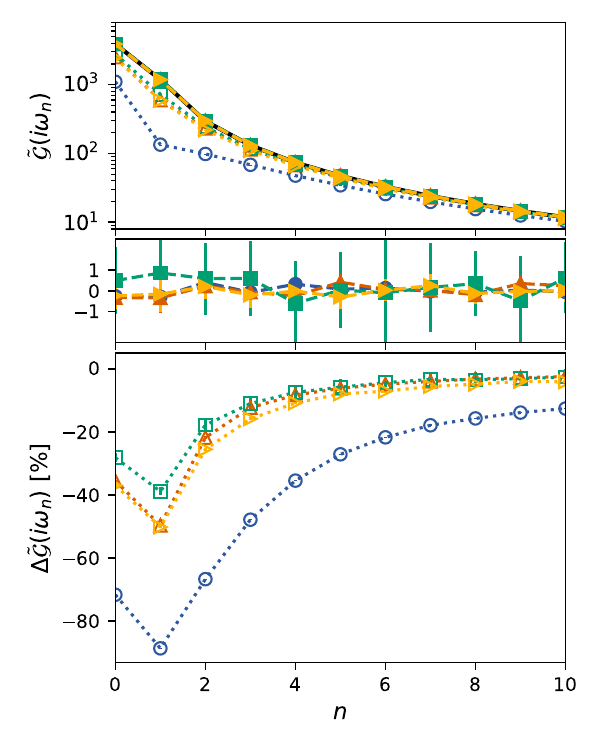}
   \caption{Plots of the periodic ($zz$) and confined ($xx$) components of the imaginary-time dipole autocorrelation function $\mathcal{\tilde G}$ in the slab geometry at variable half-thickness $R$ at $N=32$, $r_s=20$ and $\Theta=1.0$: (top left) absolute values of $\mathcal{\tilde G}(\tau)$ versus $\tau/\beta$ and (bottom left) their deviation from $\mathcal{\tilde G}^D$; (top right) the corresponding Matsubara spectrum $\mathcal{\tilde G}(\tau)$ and (bottom right) its deviation from $\mathcal{\tilde G}^D$. Statistical uncertainties are smaller than the marker size. Markers are only printed once in a few data points.
   \label{fig:slab_vs_R}
   }
\end{figure}

Figure~\ref{fig:slab_vs_R} shows clearly how the slab confinement breaks down the ideal Drude response of the electrons, $\mathcal{\tilde G}_0^D$: the free components ($zz$) of the dipole autocorrelation $\mathcal{\tilde G}(\tau)$ remain equal to $\mathcal{\tilde G}_0^D$, while the confined direction ($xx$) shows systematic deviations. From now on, we shall refrain from discussing $\mathcal{\tilde G}_{zz}$ as the differences are insignificant. In contrast, the differences $\Delta \mathcal{\tilde G}_{xx}$ are significant. In the time-domain, the differences grow toward $\tau \rightarrow \beta / 2$, whereas in the Matsubara domain $\mathcal{\tilde G}(i \omega_n)$ they die out fast with $n$.

\begin{figure}[h]
   \centering
   \includegraphics[width=6.0cm]{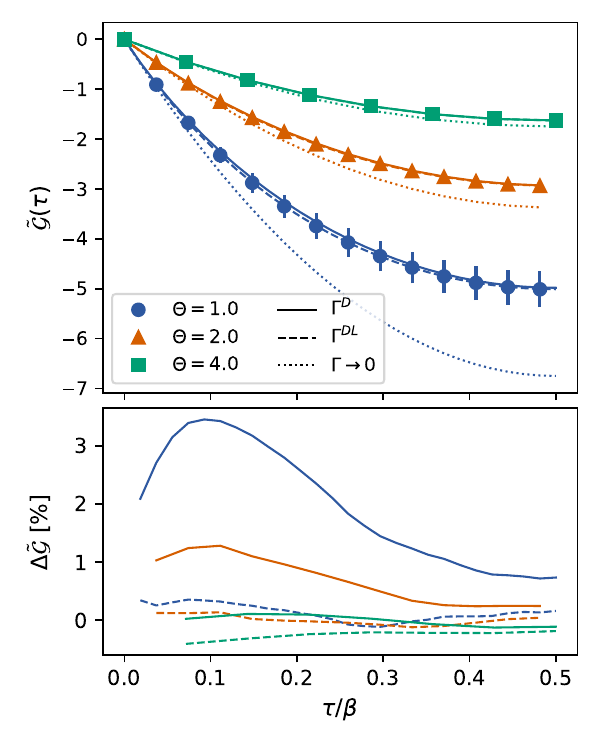}
   \caption{(top) The dipole autocorrelation function $\mathcal{\tilde G}(\tau)$ from PIMC (markers, no line) at variable temperatures with $r_s=10$, $R=10$ and $N=16$. The least-squares fitted Drude and Drude--Lorentz models $\mathcal{G}^{D}(\tau, \Gamma^D)$ (solid lines) and $\mathcal{G}^{DL}(\tau, \Gamma^{DL}, \omega_0)$ are in good agreement unlike the ideal curves $\mathcal{G}^D_0(\tau)$ (dotted lines). (bottom) Differences of the Drude (dotted lines) and Drude--Lorentz (dashed lines) models to the PIMC data.
   \label{fig:slab_Gtau_Gammafit}
   }
   
\end{figure}

To quantify this effect, we fit Drude and Drude--Lorentz models to our PIMC data as described in Sec.~\ref{sec:drude}. The fitted coefficients are collected in Table~\ref{tab:slab_Gamma_vs_rs} for $\Theta=1$ various values of $r_s$ and $R$. Similar Tables for different temperatures are provided in the SI. The fitting quality is illustrated in Fig.~\ref{fig:slab_Gtau_Gammafit}, where the apparent PIMC values of $\mathcal{\tilde G}_{xx}(\tau)$ (filled markers) at different temperatures $\Theta$ are decently captured by least-squares fitting $\Gamma^D$ (solid line) and $\Gamma^{DL}, \omega_0$ (dash-dotted line) to the data. The ideal Drude curves $\mathcal{G_0^D}$ are indicated by the dotted lines. Clearly, the Drude--Lorentz model has more expressive power than the non-resonating Drude model. However, as both parameters $\Gamma^{DL}$ and $\omega_0$ compete over a similar scattering effect, their fitting is much more prone to errors due to the statistical noise. This might call for the development of more sophisticated fitting practices and Ans\"atze. Overall, this underlines the well-known difficulty of recovering complex spectra from the imaginary-time representation \cite{Jarrell1996}.

\begin{figure}[h]
   \centering
   \includegraphics[width=6.0cm]{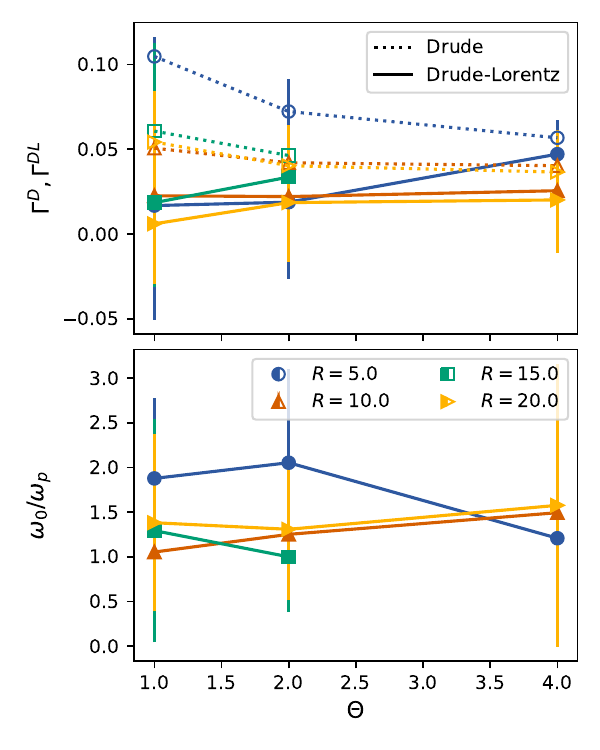}
   \caption{Least-squares fitted Drude $\Gamma^D$ and Drude--Lorentz $\Gamma^D$ (top) and $\omega_0$ (bottom) at various thicknesses versus temperature at $N=32$ at $r_s=10$.
   \label{fig:slab_Gamma_vs_theta}
   }
\end{figure}

\begin{figure}[h]
   \centering
   \includegraphics[width=6.0cm]{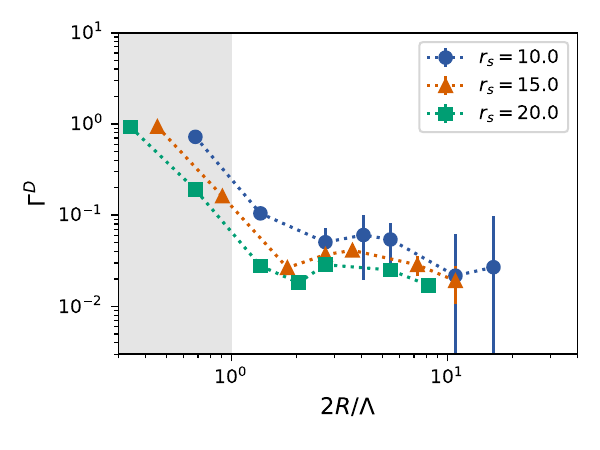}
   \includegraphics[width=6.0cm]{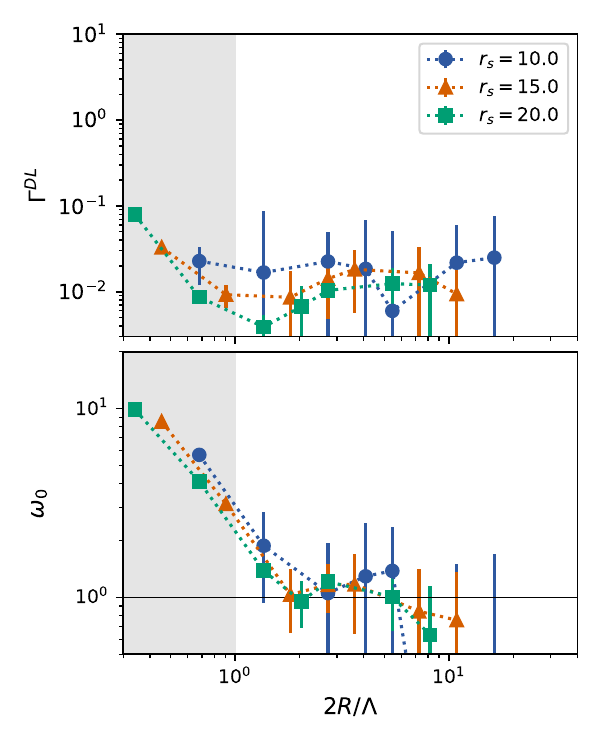}
   \caption{The effective Drude scattering $\Gamma^D$ (left) and Drude--Lorentz parameters $\Gamma^{DL}$ and $\omega_0$ versus $2R/\Lambda$ at $\Theta=1$, $N=32$. The shaded area indicates $2R < \Lambda$ regime. The values can be found in Table~\ref{tab:slab_Gamma_vs_rs}.
   \label{fig:slab_Gamma_thermal}
   }
\end{figure}

Based on the fits, we gain understanding of how the effective scattering develops with the physical parameters. Figure~\ref{fig:slab_Gamma_vs_theta} shows that $\Gamma^D$ scattering appears to decrease modestly toward higher temperatures $\Theta$. The Drude--Lorentz parameters tell a different story: the scattering rates $\Gamma^{DL}$ stay approximately constant, while the resonance $\omega_0$ undergoes a modest blue-shift. The thinnest slab $R=5$ makes an exception, because the confinement cannot so well contain the particles at lower temperatures. This is well seen in Fig~\ref{fig:slab_Gamma_thermal}: the effective scattering $\Gamma^D$ is high when $2R / \Lambda \leq 1$, where $\Lambda = \sqrt{2\lambda \beta}$ is the thermal de Broglie wavelength of the electrons. Intuitively, $2R = \Lambda$ is the threshold for a thermal electron to fit inside the potential, approximately, as a free particle without significant confinement effects. With $2R < \Lambda$, this is the main source of breakdown of the ideal Drude model, regardless of quantum many-body interactions. Furthermore, Fig.~\ref{fig:slab_Gamma_thermal} (left) also shows a steady decrease on $\Gamma^D$ toward limit $2R \gg \Lambda$, although the trend is more easily understood as that of $r_s$ versus $R$ (both $\Lambda$ and $r_s$ are constant for each individual curve). The fact that $\Gamma^D$ tends to 0 toward higher $R/r_s$ means that the thickening slab also recovers the ideal Drude response. Figure~\ref{fig:slab_Gamma_thermal} (right) shows similar effects, but the scattering from slab boundaries is not attributed to $\omega_0$, which rises sharply above $\omega_p$ when the ratio $2R / \Lambda$ decreases. When the slab grows thicker, $\omega_0$ stays close to $\omega_p$ but eventually drops. The scattering $\Gamma^{DL}$ stays lower than $\Gamma^D$ and does not vary much. Like $\Gamma^D$ it is expected to vanish slowly toward higher $R$, but the analysis is obscured by the poor signal-to-noise ratio.

\begin{figure}[h]
   \centering
   \includegraphics[width=8.0cm]{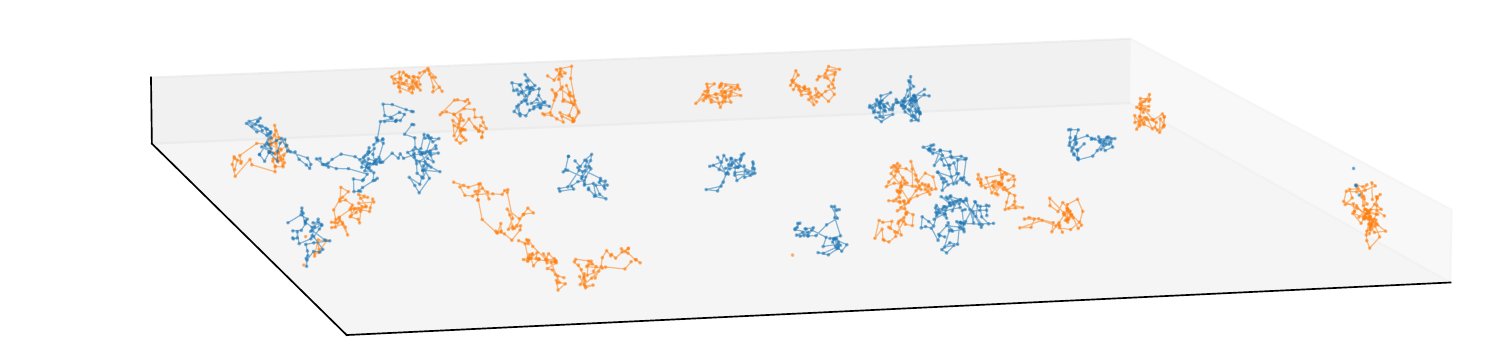}
   \includegraphics[width=8.0cm]{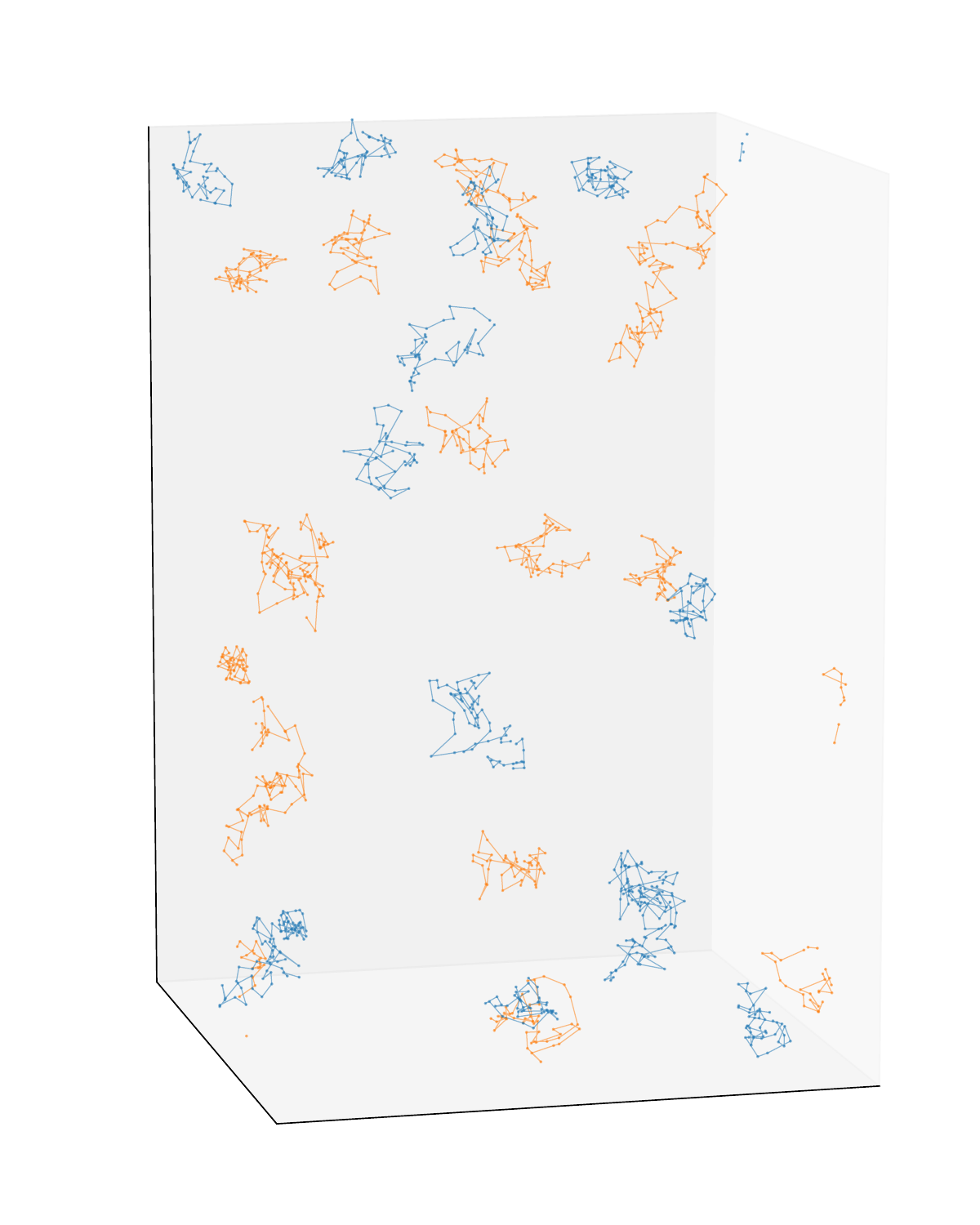}
   \caption{Snapshot particle trajectories of the slab geometry with $R=2.5$ (left) and $R=20$ (right), $r_s=6$, $\Theta=2$, $N=32$. The length scales are equal within each image but not between them. The vertical axis has open boundary conditions, while the horizontal axes are periodic. The two colors indicate different spin populations.
   \label{fig:slab_paths}
   }
\end{figure}

\begin{figure}[h]
   \centering
   \includegraphics[width=6.0cm]{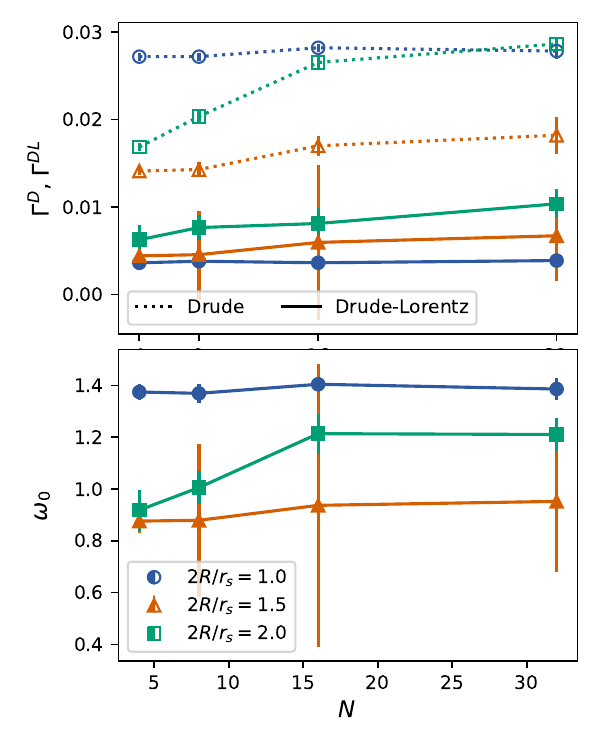}
   \caption{Fitted parameters $\Gamma^D$ and $\Gamma^{DL}$ (top) and $\omega_0$ (bottom) at $r_s=20.0$ and $\Theta=1$ using variable $N$ and $R$.
   \label{fig:slab_Gamma_vs_N}
   }
\end{figure}

The significance of exchange and Coulomb interactions may depend on the geometry: As seen in Fig.~\ref{fig:slab_paths}, slab thickness dictates the out-of-plane composition of the particles. In thin slabs ($2R/r_s \leq 1$; Fig.~\ref{fig:slab_paths} top) the particles assume a one-layer compositions. Then, many-body interactions along the out-of-plane direction only occur through in-plane distortions and are generally modest. In contrast, thicker slabs ($2R/r_s \gg 1$; Fig.~\ref{fig:slab_paths} bottom) enable vertical stacking to multiple layers, between which the many-body effects grow more significant. This is exactly seen in Figure~\ref{fig:slab_Gamma_vs_N}, where the particle number $N$ is used to gauge the many-body correlation effects in the fitted Drude (dotted lines) and Drude--Lorentz (solid lines) parameters toward the thermodynamic limit $N\rightarrow \infty$. The thin, planar stacking $2R/r_s \leq 1$ (blue curves) shows no noticeable dependence on $N$ for any of the parameters. As the ratio grows above 1 (green and orange curves), the particles stack up show increasing sensitivities to $N$. Here, the effect is accentuated by the truncation of periodic images (see Sec.~\ref{sec:computational_details}). The Drude--Lorentz model reveals that this effect is more pronounced in the $\omega_0$ resonance, while the Drude model only shows it in $\Gamma^D$. Crossing of the curves indicates that the many-body interaction may have significant effects on the relative scattering rates. In general, this also calls for careful consideration of the finite-size simulation effects, but will not study them further in this work.

\begin{figure}[h]
   \centering
   \includegraphics[width=6.0cm]{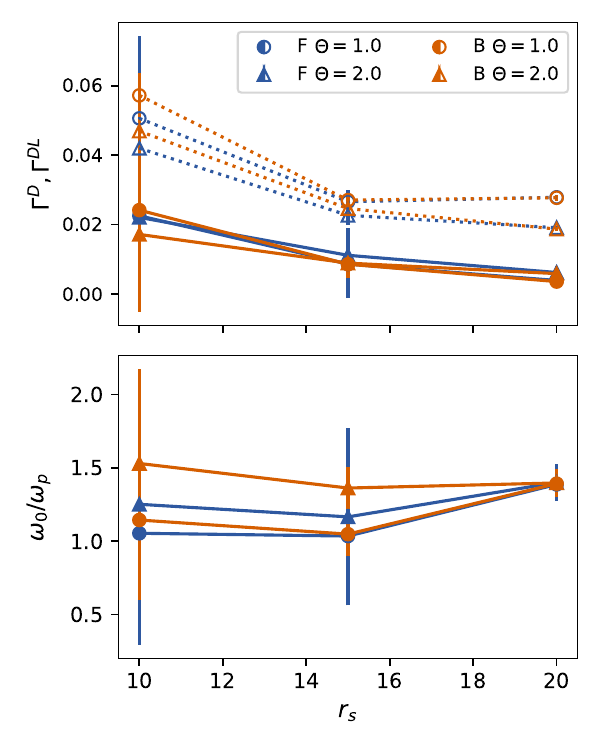}
   \caption{Dependence on the quantum statistics: Least-squares fitted Drude and Drude--Lorentz parameters between Fermion (F; dashed lines) and Boltzmannon (B; dotted lines) simulations of $R=10$, $N=32$. The lines are drawn to guide the eyes.
   }
   \label{fig:slab_vs_kind}
\end{figure}

Finally, in Fig.~\ref{fig:slab_vs_kind} we plot how quantum statistics affects the scattering properties. The blue curves with dotted lines are based on proper Fermion simulations, whereas the orange curves with dashed lines are based on distinguishable Boltzmannon simulations. At low densities $r_s \geq 15$ the data are identical but at $r_s=10$ the Fermion scattering is systematically lower, if only modestly. The effect is expected and likely more intense toward lower $r_s$ and $\Theta$. This is also where the average sign degrades rapidly. Thus, this aspect is only considered lightly to illustrate its existence. An explanation is that the exchange hole between identical electrons suppresses the scattering via Coulomb interaction.

\begin{table}[h]
    \centering
    \begin{tabular*}{\linewidth}{@{\extracolsep{\fill}} cccr|cccc}
$r_s$       &   $\omega_p$ &          $R$ & $2R / \Lambda$ & $\langle s \rangle'$ &   $\Gamma^D$ & $\Gamma^{{DL}}$ & $\omega_0 / \omega_p $  \\ \hline
6.00        &       0.1179 &          2.5 &        1.128 &      0.25(5) &       0.3(2) &       0.3(3) &         1(3)  \\
            &              &          5.0 &        2.256 &     0.028(8) &              &              &               \\
            &              &         10.0 &        4.511 &     0.021(8) &              &              &               \\
            &              &         15.0 &        6.767 &     0.009(6) &              &              &               \\
            &              &         20.0 &        9.022 &     0.007(6) &              &              &               \\
            &              &         40.0 &       18.044 &     0.005(8) &              &              &               \\
            &              &         60.0 &       27.067 &    -0.002(7) &              &              &               \\
10.00       &       0.0548 &          2.5 &        0.680 &     0.925(6) &    0.722(10) &    0.023(13) &      5.69(9)  \\
            &              &          5.0 &        1.361 &      0.51(2) &    0.105(14) &      0.02(8) &      1.9(10)  \\
            &              &         10.0 &        2.722 &    0.206(15) &      0.05(3) &      0.02(3) &       1.1(9)  \\
            &              &         15.0 &        4.082 &      0.17(2) &      0.06(6) &      0.02(6) &      1.3(10)  \\
            &              &         20.0 &        5.443 &     0.109(8) &      0.05(4) &      0.01(4) &       1.4(9)  \\
            &              &         40.0 &       10.887 &    0.090(14) &      0.02(4) &      0.02(5) &      0.0(14)  \\
            &              &         60.0 &       16.330 &    0.078(13) &              &      0.02(5) &      0.3(14)  \\
15.00       &       0.0298 &          2.5 &        0.454 &     0.993(3) &     0.934(7) &     0.033(5) &      8.54(7)  \\
            &              &          5.0 &        0.907 &     0.960(3) &     0.162(3) &     0.009(3) &      3.13(5)  \\
            &              &         10.0 &        1.814 &    0.545(10) &     0.027(3) &     0.009(8) &       1.0(3)  \\
            &              &         15.0 &        2.722 &    0.642(13) &     0.037(6) &    0.014(10) &       1.2(4)  \\
            &              &         20.0 &        3.629 &    0.607(14) &     0.041(5) &    0.018(13) &       1.2(5)  \\
            &              &         40.0 &        7.258 &    0.312(15) &     0.029(8) &      0.02(2) &       0.8(7)  \\
            &              &         60.0 &       10.887 &      0.30(2) &    0.019(13) &    0.009(11) &       0.8(8)  \\
20.00       &       0.0194 &          2.5 &        0.340 &          1.0 &     0.928(9) &     0.079(5) &      9.87(8)  \\
            &              &          5.0 &        0.680 &   0.9957(10) &   0.1904(15) &     0.009(2) &      4.11(4)  \\
            &              &         10.0 &        1.361 &     0.946(4) &    0.0278(8) &   0.0039(14) &      1.39(6)  \\
            &              &         15.0 &        2.041 &      0.69(2) &     0.018(2) &     0.007(5) &       1.0(3)  \\
            &              &         20.0 &        2.722 &     0.941(4) &    0.0287(8) &     0.010(2) &      1.21(7)  \\
            &              &         40.0 &        5.443 &    0.618(13) &     0.025(3) &     0.012(7) &       1.0(3)  \\
            &              &         60.0 &        8.165 &      0.50(3) &     0.017(5) &     0.012(9) &       0.6(5)  \\
    \end{tabular*}
    \caption{Fitted Drude and DL parameters, average signs and derived quantities of PIMC simulations at $\Theta=1$, $N=32$ and variable $r_s$ and $R$. Statistical uncertainties are given in parentheses, and the poorest of fits are not printed.
    \label{tab:slab_Gamma_vs_rs}
    }
\end{table}

\begin{table}[h]
    \centering
    \begin{tabular*}{\linewidth}{@{\extracolsep{\fill}} ccc|ccc}
$R$         &     $\Theta$ &    $\Lambda$ & $\Gamma^D$   & $\Gamma^{DL}$ & $\omega_0 / \omega_p $ \  \\ \hline
5.0         &          1.0 &         7.35 &    0.105(13) &      0.02(8) &      1.9(10)  \\
            &          2.0 &         5.20 &      0.07(2) &      0.02(5) &      2.1(10)  \\
            &          4.0 &         3.74 &     0.057(7) &      0.05(2) &      1.2(12)  \\
10.0        &          1.0 &         7.35 &      0.05(2) &      0.02(3) &      1.1(10)  \\
            &          2.0 &         5.20 &    0.042(11) &      0.02(2) &       1.3(6)  \\
            &          4.0 &         3.74 &    0.040(11) &      0.03(3) &      1.5(13)  \\
15.0        &          1.0 &         7.35 &      0.06(4) &      0.02(7) &      1.3(13)  \\
            &          2.0 &         5.20 &    0.046(15) &      0.03(2) &       1.0(8)  \\
20.0        &          1.0 &         7.35 &      0.05(4) &      0.01(5) &      1.4(10)  \\
            &          2.0 &         5.20 &      0.04(3) &      0.02(4) &      1.3(11)  \\
            &          4.0 &         3.74 &      0.04(3) &      0.02(4) &         2(2)  \\
    \end{tabular*}
    \caption{Fitted parameters of the Drude and DL model at various temperatures based on $N=32$ and $r_s=10$ with $\omega_p=0.05478$.
    \label{tab:slab_Gamma_vs_theta}
    }
\end{table}

\subsection{Summary and outlook}
\label{sec:summary}

We have demonstrated a PIMC method for studying the optical response properties of the HEG confined to a finite slab. We focus on the breakdown of the ideal Drude response arising from the confinement. In principle, our method addresses these effects accurately at finite densities and temperatures, producing benchmark data to inform further applications and the development of other numerical approaches. In practice, limitations ensue because of the well-known FSP, which undermines the numerical efficiency toward larger sizes and lower temperatures. In this proof-of-concept study we focused in providing a high throughput of results on selected intuitive principles, leaving the scrutiny of chasing perfect accuracy, computational barriers or specific applications to another time.

We show that the slab confinement causes a finite Drude scattering for the out-of-plane optical susceptibility, whereas the in-plane response remains unaffected or very modestly changed. The scattering rate is treated by fitting to the phenomenological Drude model, which allows the most straightforward analytic continuation of the complex response. The more versatile Drude--Lorentz model is also studied, but more development and lower uncertainties are needed to disentangle its spectral features in high confidence. This underlines the infamous difficulty of informing sophisticated spectra with noisy imaginary-time data, but despite the challenges, this remains an intriguing research front.

Judging by the Drude scattering, certain trends with the temperature are apparent: When the confinement size drops below the thermal wavelength, surface effects dominate the scattering response. When the confinement size is large, the scattering tends asymptotically toward the zero limit, \textit{i.e.}, bulk response. In between, when the geometry accommodates a few out-of-plane layers of electrons, quantum many-body interactions play a role in the collective scattering. The different scattering channels are on equal footings and cannot be singled out. When the particle density is high enough, the Fermion exchange grows important by lowering the scattering compared to a non-permuting reference. We believe that similar principles manifest also in resembling confinements, such as cylindrical nanorods, nanospheres or less ideal atomistic potentials \cite{Povarnitsyn2012}. Overall, it is clear that quantum many-body effects cannot be overlooked when treating the optical response of electron plasma confined to nanometer scales.

\subsection{Acknowledgments}
\label{sec:acknowledgements}
The authors wish to acknowledge CSC – IT Center for Science, Finland, and the Tampere Center for Scientific Computing; for computational resources. The authors acknowledge the financial support from the Photonics Research and Innovation Flagship (PREIN - decision 320165) and the Research Council of Finland project AQUA-PHOT (decision Grant No. 349350). JT acknowledges Ilkka Kylänpää for his support in the implementation of permutation sampling.

\bibliographystyle{ieeetr}
\bibliography{references}
\end{document}